\documentclass[journal, two side]{IEEEtran}
\ifCLASSINFOpdf
\else
\fi

\usepackage{hyperref}
\usepackage{graphicx}
\usepackage{color}
\DeclareGraphicsExtensions{.pdf,.jpeg,.png}

\definecolor{violet}{rgb}{0.75,0,1}

\begin{document}
%
\title{On Using Complex Event Processing for Dynamic Demand Response Optimization in Microgrid}



\author{\IEEEauthorblockN{
Qunzhi Zhou$^*$, Yogesh Simmhan$^{\S}$ and Viktor Prasanna$^{\S}$ \\
\IEEEauthorblockA{* Department of Computer Science, University of Southern California, USA\\
Email: qunzhizh@usc.edu\\
$^{\S}$ Ming Hsieh Department of Electrical Engineering, University of Southern California, USA\\
Email: \{simmhan, prasanna\}@usc.edu}\\
}
}



%


\markboth{Proceedings of Green Energy and Systems Conference 2013, November 25, Long Beach, CA, USA. }{This full text paper was peer reviewed at the direction of Green Energy and Systems Conference subject matter experts.}

\maketitle

\begin{abstract}
Demand-side load reduction is a key benefit of Smart Grids. However, existing demand response
optimization (DR) programs fail to effectively leverage the near-realtime information available from
smart meters and Building Area Networks to respond dynamically to changing energy use
profiles. We investigate the use of semantic Complex Event Processing (CEP) patterns to model and detect
dynamic situations in a campus microgrid to facilitate adaptive DR. Our focus is on demand-side
management rather than supply-side constraints. Continuous data from information sources like
smart meters and building sensors are abstracted as event streams. Event patterns
for situations that assist with DR are detected from them. Specifically, we offer a
taxonomy of event patterns that can guide operators to define situations of
interest and we illustrate its efficacy for DR by applying these patterns on realtime events in the USC Campus
microgrid using our CEP framework.

\end{abstract}


%
\IEEEpeerreviewmaketitle

\section{Introduction}
Smart Grids provide realtime monitoring capability of interconnected power grid elements, two-way
communications between end-use devices, customers and utilities, and the opportunity to integrate
and use information from diverse sources such as weather forecasts and event schedules. This
information infrastructure enables the design of advanced information technology systems to
improve the power grid efficiency and meet the rapidly increasing electricity demand.


Demand response optimization (DR) is a cornerstone component of Smart Grids, and deals with
managing demand-side load in response to supply conditions. 
Traditional DR approaches require advanced planning, hours or days ahead, and operate on a broadcast
principle that reaches to all customers. As the energy usage patterns of customers change, a more
effective strategy is to target the most relevant 
customers or loads based on the current or impending energy usage profiles to meet the required
curtailment target. The notion of \emph{dynamic DR} (D2R) uses near-realtime information to
understand dynamic energy consumption situations, and responds with precise curtailment actions, with
low latency and high relevance. In particular, we focus on curtailing demand-side consumption when
the supply-side constraints are known, using the USC Campus as a microgrid testbed to evaluate our
approaches.

Complex Event Processing (CEP) is an
information processing framework to detect the occurrence of specified pattern of events by examining hundreds or
thousands of event data streams with a low latency on the order of seconds. 
CEP can help correlate continuous streams of data from the smart grid, and perform online analysis
to detect situations of interest modeled as event patterns. For e.g., a CEP pattern could detect an
opportunity for temperature reset in a classroom if it is not occupied, the setpoint temperature for
the room is less than 72$^{\circ}$F and no classes are scheduled for the next hour. Such insight
into ongoing situations enables timely and opportunistic DR curtailment responses. While the
application of CEP to smart grids is innovative, existing literature just offers anecdotal
applications of CEP to smart grids and are tightly scoped to narrow scenarios.  In particular, there
is a lack of a detailed exploration into the categories of CEP 
patterns that can benefit demand-side management in DR, an accessible means to define them at a
higher level of abstraction, and a practical illustration of such patterns in action. Such an effort
will both convince operators of the novel value of defining CEP patterns and also ease the process of
defining the patterns using exemplar pattern templates.

In this paper, we make the following contributions, 

\begin{enumerate}
    \item We introduce the use of semantic CEP framework to model event patterns relevant to DR (\S~\ref{sec:background}).
    \item We discuss a taxonomy of event patterns to guide different aspects of DR, along with
      examples for demand management in the USC campus microgrid (\S~\ref{sec:taxonomy}).
    \item We evaluate the efficacy of event-based DR by presenting pattern detection statistics
      from USC campus microgrid experiments(\S~\ref{sec:casestudy}).
\end{enumerate}


\section{Approach Overview}
\label{sec:background}
\subsection{USC Campus Microgrid}
The University of Southern California serves as a testbed to experiment
and evaluate DR technologies as part of the Department of Energy-sponsored Los Angeles Smart Grid Demonstration Project \cite{YogeshReport2011}.
In particular, the research focus is on demand-side load management, under the assumption that the
supply-side characteristics are known.
USC 
is the largest private customer of the Los Angeles Department of Water and Power (LADWP) with over
60,000 
students, faculty and staff spread over 170 buildings containing classrooms, residence halls,
offices, labs, hospitals and restaurants. USC is well instrumented, with a campus-wide Building Area
Network (BAN) that can monitor buildings and equipment to measure power usage, operational status, space
and setpoint temperatures, airflow, occupancy and so on at minute-intervals from the Campus Energy
Control Center.

The demand curtailment strategies available within the campus microgrid include direct control strategies, like Global Temperature
Reset (GTR) and Duty Cycling, and voluntary curtailment strategies through email notifications
sent to building occupants. 
Currently, these strategies are scheduled at pre-determined time periods, day(s) ahead, based on historical
power usage trends.
However, these strategies need to be initiated based on near-realtime energy usage conditions
and also supplemented with nimble strategies that leverage dynamic demand reduction opportunities on campus.
Realtime buildings and equipment information from
the campus BAN, campus schedule, facility details and weather forecasts
can be analyzed to detect additional curtailment opportunities.

\subsection{Semantic Complex Event Processing}
CEP is an information processing framework to detect the occurrence of
a pattern of events over event streams.
Continuous, time-series data from sensors and other information sources in the microgrid can be
abstracted as event streams. For example, an event stream may be comprised of timestamped KWh energy usage
of an HVAC (heating, ventilation, and air conditioning) unit in a particular room reported every minute. Weather conditions for a
particular zipcode provided every hour by the NOAA (National Oceanic and Atmospheric Administration) web service can also form an event stream. 
Dynamic DR situations of interest are modeled as combination of these event occurrences, i.e., as event
patterns.

\begin{figure}[!h!t!b]
  \centering
  \includegraphics[width=3.3in]{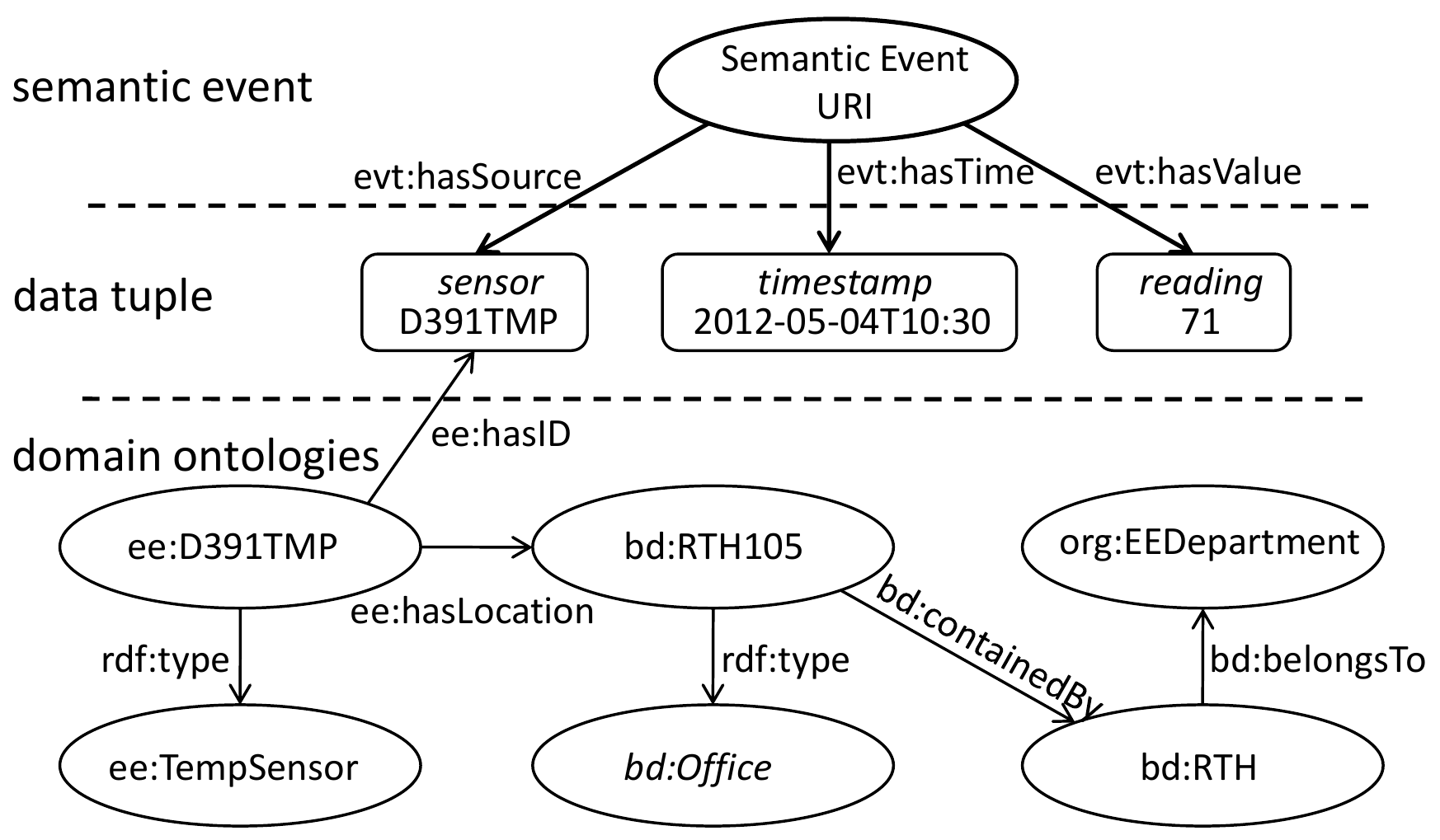}
  \caption{Semantic Event for Space Temperature Stream}
  \label{fig:sevent}
\end{figure}

Manually defining fine-grained patterns over events can be tedious, especially in an information rich
domain such as a microgrid. Our semantic CEP framework, \emph{SCEPter} \cite{zhou2012SCEPTechReport,
  zhou2012ISWC}, facilitates DR event pattern specification using domain concepts. This
makes it more user-friendly compared to existing
CEP systems that process events as relational data tuples and require precise knowledge 
of the raw event. 
Our semantic CEP framework 
incorporates domain knowledge by using semantic ontologies \cite{zhou2012ITNG}. Figure
\ref{fig:sevent} is the semantic graph model of a raw event data tuple, showing how the concepts and
entities in the microgrid are related, such as facilities (bd:RTH105), their types (bd:Office),
sensors (ee:D391TEMP), measured variables (ee:TempSensor), and organizations
(org:EEDepartment). Users can specify patterns at the domain level without knowing the raw event
details.

Patterns over semantic events are specified using a two-segment pipeline query model, i.e.,
\newline
\\\hspace*{0.2cm}\texttt{\small Semantic Event Pattern ::=}
\\ \hspace*{0.8cm}\texttt{\small output definition}
\\ \hspace*{0.8cm}\texttt{\small stream declaration}
\\ \hspace*{0.8cm}\texttt{\small [semantic filtering subpattern]* |}
\\ \hspace*{0.8cm}\texttt{\small [syntactic CEP subpattern]?}
\newline
\newline
\noindent where the output definition projects event attributes to query results using keyword \emph{SELECT}, stream declaration associates event variables with source streams using keyword \emph{FROM}. Semantic filtering
subpatterns are represented using SPARQL \cite{website:Sparql2} queries and CEP subpatterns are
represented using event processing language provided by the native CEP engine. For example, the
following pattern specifies the temperature readings in an \emph{office} room increased by
\emph{3$^\circ$F} within 5 minutes on the space temperature stream \emph{stsstream}.
\newline
\newline
\noindent\hspace*{0.2cm}\texttt{\small SELECT(?e1,?e2) FROM(?e1,?e2,stsstream)}
\\ \hspace*{0.2cm}\texttt{\small WHERE \{?e1 evt:hasSource ?src . ?e2 evt:}
\\ \hspace*{0.2cm}\texttt{\small hasSource ?src . ?src bd:hasLocation ?loc . }
\\ \hspace*{0.2cm}\texttt{\small ?loc rdf:type bd:Office\} | SEQ(?e1, }
\\ \hspace*{0.2cm}\texttt{\small ?e2(reading-?e1.reading>3) within 5min)}
\newline
\newline
\noindent Here \emph{office} is a conceptual term, available to the user, which transparently maps to rooms classified
as such within the ontology. \emph{SEQ} is an event processing operator to correlate sequential
events. Other operators also include \emph{JOIN} correlation as well as aggregation operators \emph{SUM}, \emph{AVG}.

More advanced semantic CEP patterns, discussed later, can be used to correlate events from multiple streams to
detect meaningful DR situations for realtime decision support. 

\subsection{System Architecture}

\begin{figure}[!h!t!b]
  \centering
  \includegraphics[width=3.3in]{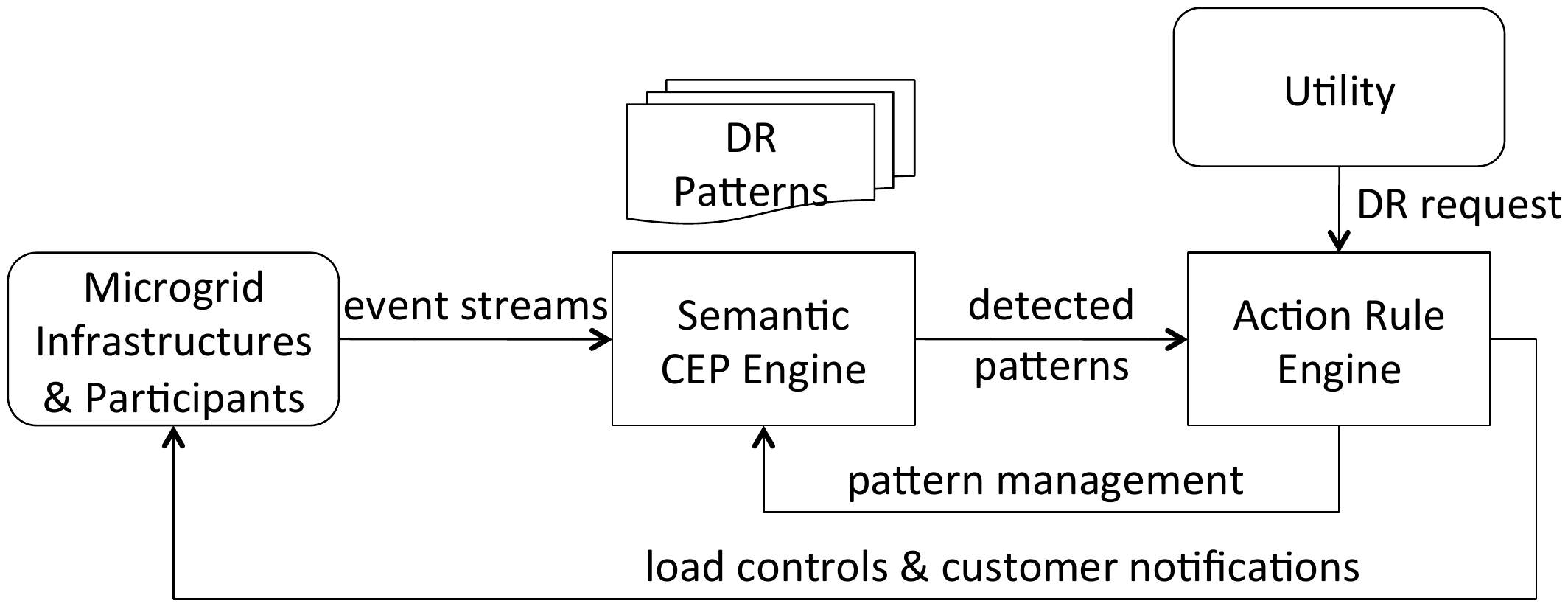}
  \caption{Microgrid Event-based D2R Architecture}
  \label{fig:ra}
\end{figure}
Figure \ref{fig:ra} shows the proposed architecture of event-based D2R in the campus microgrid. Our semantic
CEP engine matches patterns by monitoring event streams from sensors in the microgrid. Another part of the architecture, i.e., the action rule engine, which map detected patterns to curtailment actions is currently
under development. 
Pattern actions include direct control of equipment, such as GTR and duty cycling, in buildings
determined by the pattern, sending notifications to DR participants, or activating additional patterns. In the following we focus on the core event processing engine and discuss event
patterns that can be defined for dynamic DR optimization in microgrid.

\section{DR Event Pattern Taxonomy}
\label{sec:taxonomy}
\begin{figure}[!h!t!b]
  \centering
  \includegraphics[width=3.5in]{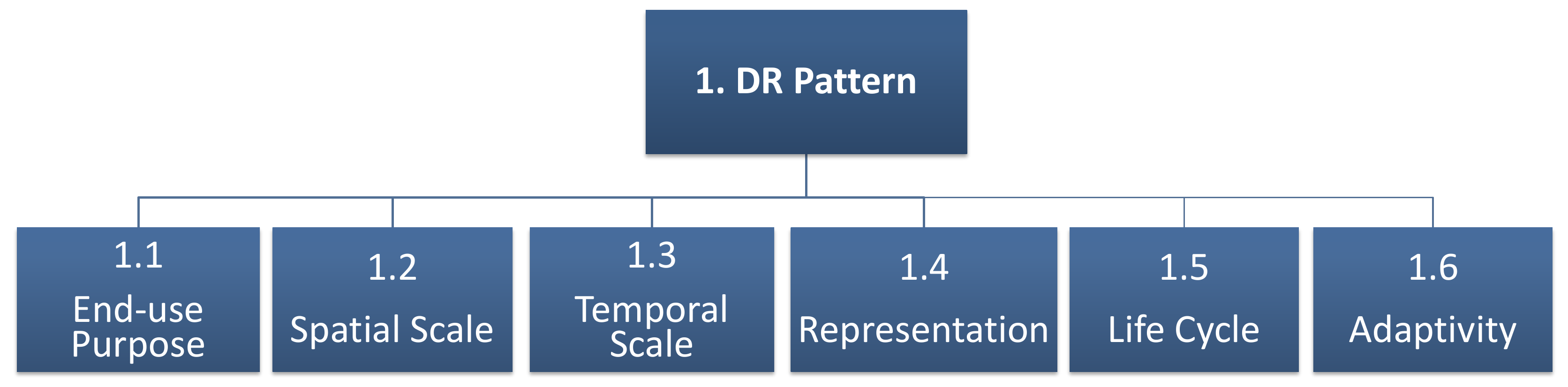}
  \caption{Top-level \emph{orthogonal} dimensions of DR Pattern Taxonomy}
  \label{fig:taxonomy1}
\end{figure}

The potential space of event patterns is enormous. In the absence of investigation and
classification, it becomes onerous  
for operators to go beyond facile patterns and exploit the innovation and expressivity of semantic CEP patterns for different
aspects of dynamic DR. We offer a taxonomy of DR event patterns motivated from scenarios and semantic concepts
observed in the USC microgrid, but generalizable to other environments. Figure \ref{fig:taxonomy1}
shows top-level \emph{orthogonal dimensions} of this taxonomy that are key characteristics to
consider when defining a DR pattern. Patterns are not exclusive to one dimension
but have a specific feature from each dimension.

\subsection{End-use Purpose Dimension}
\begin{figure}[!h!t!b]
  \centering
  \includegraphics[width=2.8in]{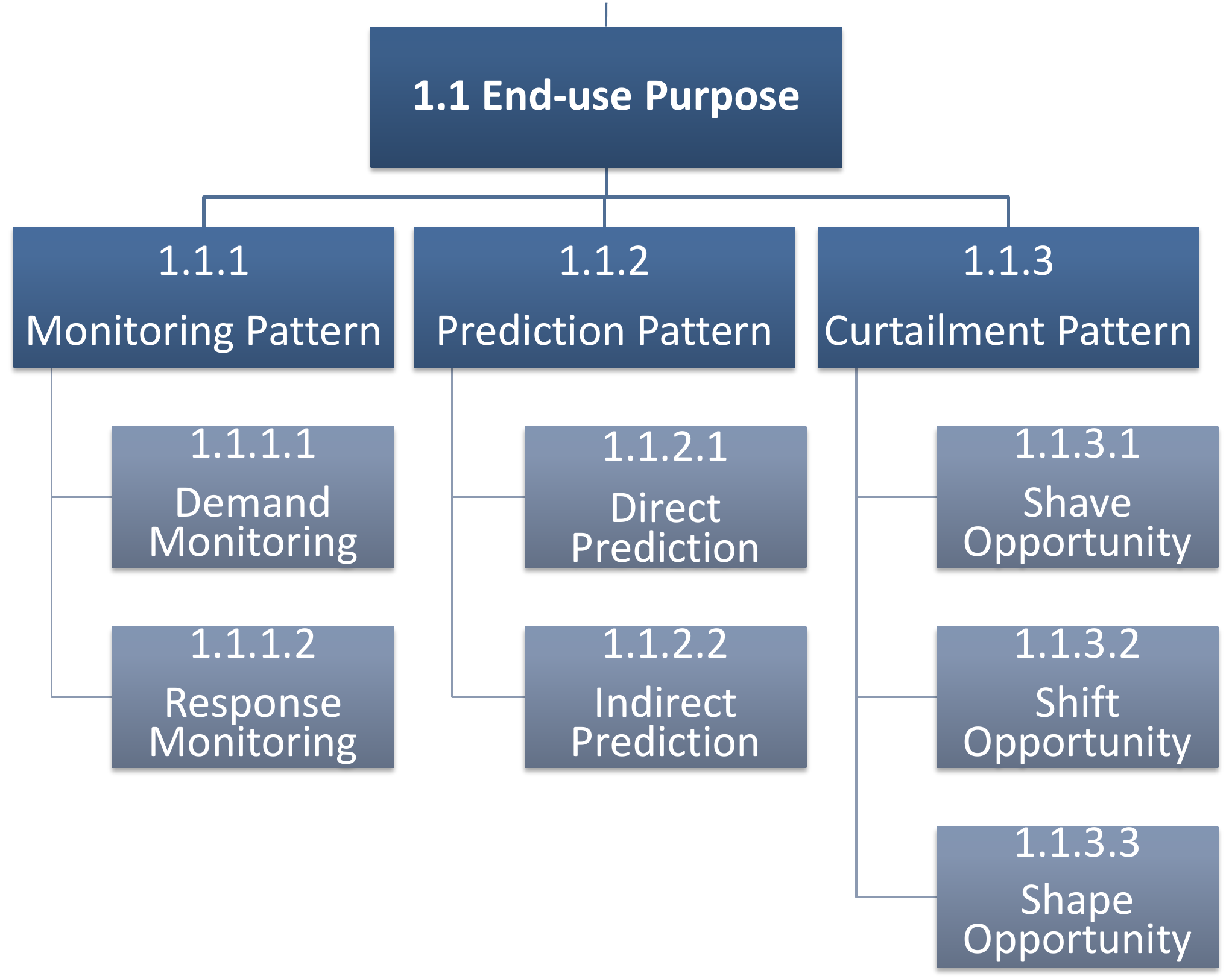}
  \caption{End-use Purpose dimensions of DR Pattern Taxonomy}
  \label{fig:taxonomy2}
\end{figure}

Patterns can be categorized based on the objective of their end-use, as shown in Figure
\ref{fig:taxonomy2}. These categories are typically exclusive. The obvious example are curtailment
patterns that can identify curtailment opportunities that can detect transient power wastage, or
trigger direct and voluntary curtailment actions.  However, patterns can also play a role in
situation monitoring and early warning. Meter readings can be aggregated to monitor demand levels,
and indirect influencers of power usage used to predict demand trends. These monitoring and
prediction patterns can trigger control/notification actions or initiate detection of specific
curtailment patterns. This can enable incremental and opportunistic DR curtailment.

\subsubsection{Monitoring Pattern}
Patterns in this category evaluate demand profiles of spaces and equipments at fine granularity by
analyzing and aggregating meter and sensor data. A sample monitoring pattern is,

\textbf{Example 1.} The power used by building \emph{``MHP''}, \emph{averaged} over 5 minutes,
exceeds a given pre-peak load of 27~KW.

Let \emph{?m} represents events from the meter's KW reading stream \emph{meterstream}, the pattern to
detect the above situation is,
\newline
\newline
\noindent\hspace*{0.5cm}\texttt{\small SELECT(avg) FROM(?m,meterstream)}
\\ \hspace*{0.5cm}\texttt{\small WHERE \{?m evt:hasSource ?src . ?src }
\\ \hspace*{0.5cm}\texttt{\small bd:hasLocation bd:MHP\} | AVG(?m) AS avg }
\\ \hspace*{0.5cm}\texttt{\small WINDOW(?m,sliding,5min) HAVING(avg>27)}
\\

\noindent The above pattern detects the need for curtailment and help initiate low-latency
curtailment strategies such as changing the setpoint of a variable frequency drive unit in the
building where the pattern was seen to avoid peak demand.

We further classify monitoring patterns as demand monitoring and response monitoring patterns
(Figure \ref{fig:taxonomy2}). Example 1 is a demand monitoring pattern. Response monitoring patterns
evaluate the effectiveness of a curtailment operation, and can be used to determine if a more
aggressive curtailment strategy is required. An example response monitoring situation is,

\textbf{Example 2.} 15 minutes \emph{after} a global temperature reset (GTR) operation was performed in \emph{``MHP''}, the building's power consumption remains greater than 30~KW.

A sequence CEP pattern can be used to detect such an insufficient curtailment situation and trigger
further actions such as HVAC unit duty cycling.



\subsubsection{Prediction Pattern}
Traditional demand prediction models are ill suited for energy forecast at fine temporal and
spatial scales, particularly as consumption profiles change~\cite{Bhattacharyya2009}. 
In a campus microgrid, dynamic events like scheduling or cancellation of
classes, space occupancy changes and holidays can help predict power consumption trends~\cite{SaimaDDDM}.
Prediction patterns are categorized as direct and indirect
predictions (Figure \ref{fig:taxonomy2}). Direct predictions forecast demand solely based on prior
energy consumption using timeseries models or historical baselines. 
Alternatively, indirect predictions combine demand influencers to predict future changes in demand.

\textbf{Example 3.} Power usage in an empty computer lab is currently $<$ 0.5KW, and a class is scheduled in 1~hour.

Semantic subpatterns can be defined over class schedule stream \emph{schstream} and meter measurement stream to filter events based on the location type. Qualified events, denoted as \emph{?m} and \emph{?c}, can be piped to the following \emph{JOIN} CEP subpattern,
\newline
\newline
\noindent\hspace*{0.5cm}\texttt{\small JOIN(?m,?c) ON(?c.schedule-?m.timestamp }
\\ \hspace*{0.5cm}\texttt{\small <3600,?m.reading<0.5)}
\newline

\subsubsection{Curtailment Pattern}
Curtailment patterns identify distributed and dynamic curtailment opportunities which supplement
scheduled or voluntary curtailments. These patterns can be defined by DR participants ranging from
facility managers to department coordinators to end users.

Curtailment patterns can be
classified based on the action taken as shave, shift or shape (Figure \ref{fig:taxonomy2}). Shave
patterns detect non-critical or wasteful power
usage that can be eliminated.

\textbf{Example 4.} The temperature in a meeting room is lower than 73~$^\circ$F when it is unoccupied.
%

Shift patterns identify non-urgent power demand from certain equipments which can be rescheduled to
off-peak periods. Such equipments may include washing machines and campus EVs. Lastly, shaping
patterns flatten demand curves by dynamically selecting, for example, HVAC units to duty cycle:

\textbf{Example 5.} More than 6 fan coils are operating concurrently in \emph{``MHP''} during peak hours.


\subsection{Spatial Scale}
\begin{figure}[!h!t!b]
  \centering
  \includegraphics[width=3.5in]{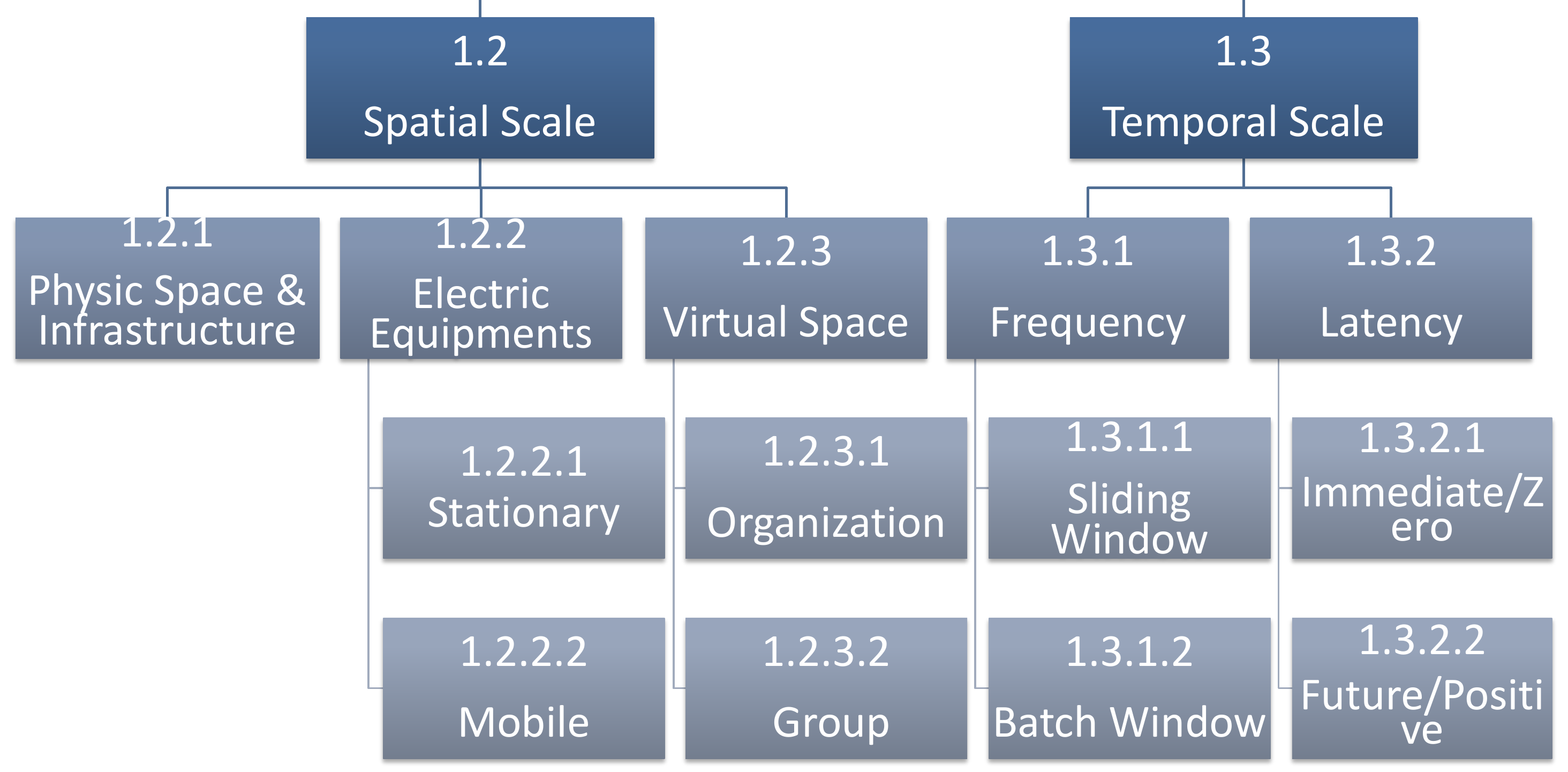}
  \caption{Spatial and Temporal dimensions}
  \label{fig:taxonomy3}
\end{figure}

Besides end-use purpose, DR patterns are usually associated with a spatial dimension. This dimension
helps identify the target spatial entity, either physical or virtual, on which some end-use action is
required (Figure \ref{fig:taxonomy3}).

\subsubsection{Physical Space and Equipment}

Physical grid entities include campus, buildings, rooms as well as individual
equipment. 
The spatial granularity may vary by  DR participants or end use. For example, campus
managers can specify campus-level monitoring patterns to trigger global curtailment operations, while building managers define
room or equipment demand prediction and curtailment patterns.
Physical objects can be further classified as stationary and mobile. The latter include EVs and
portable appliances and benefit from the transparency offered by semantic patterns in masking
variation in their physical event streams based on their location.

%
%
\subsubsection{Virtual Space}
DR Patterns can also be defined for virtual spaces or objects such as organizations and customer
segments. Virtual spaces may be physically contiguous, such as a department located in
neighboring buildings, or scattered, such as a customer segment that is environmentally
conscious.

\textbf{Example 6.} The total power demands from \emph{EE department} exceed 600~KW.\\

\noindent\hspace*{0.2cm}\texttt{\small SELECT(sum) FROM(?m, meterstream)}
\\ \hspace*{0.2cm}\texttt{\small WHERE \{?m evt:hasSource ?src . ?src }
\\ \hspace*{0.2cm}\texttt{\small bd:hasLocation ?loc . ?loc bd:belongsTo }
\\ \hspace*{0.2cm}\texttt{\small org:EEDepartment\} | SUM(?m) AS sum}
\\ \hspace*{0.2cm}\texttt{\small HAVING(sum>600)}
\\

\noindent Upon detecting the above pattern over the (virtual) department space, the department's
coordinator can be notified to initiate local curtailment strategies within the department.

%
%
%

\subsection{Temporal Scale}
The interval nature of events means that DR pattern include temporal properties such as the
frequency of evaluating patterns and the latency time for response after detection (Figure \ref{fig:taxonomy3}).

\subsubsection{Frequency}
The frequency of a DR pattern is determined by its time window constraints. There are two types of
time windows: sliding and batch. Window width can either be specified using the number of events or a
length of time period. For a sliding window, events are processed by gradually moving the window in
single event increments. Example 1 used a sliding window. For a batch window, events are processed by moving the window in discrete,
non-overlapping time/event blocks. 
A batch window is useful, for example, when we
want to monitor a building's aggregated consumption every hour.

%
%

\subsubsection{Latency}
The latency of a pattern is the difference between the time of its detection and the time of its
consequence. Most patterns, including the monitoring and curtailment patterns, have immediate impact,
i.e., zero latency. A prediction pattern however has a positive latency as it is anticipatory and detects a future
situation. A curtailment pattern may also have a positive latency when it is used to schedule a
future curtailment operation rather than trigger one immediately.


\subsection{Representation}
\begin{figure}[!h!t!b]
  \centering
  \includegraphics[width=3.4in]{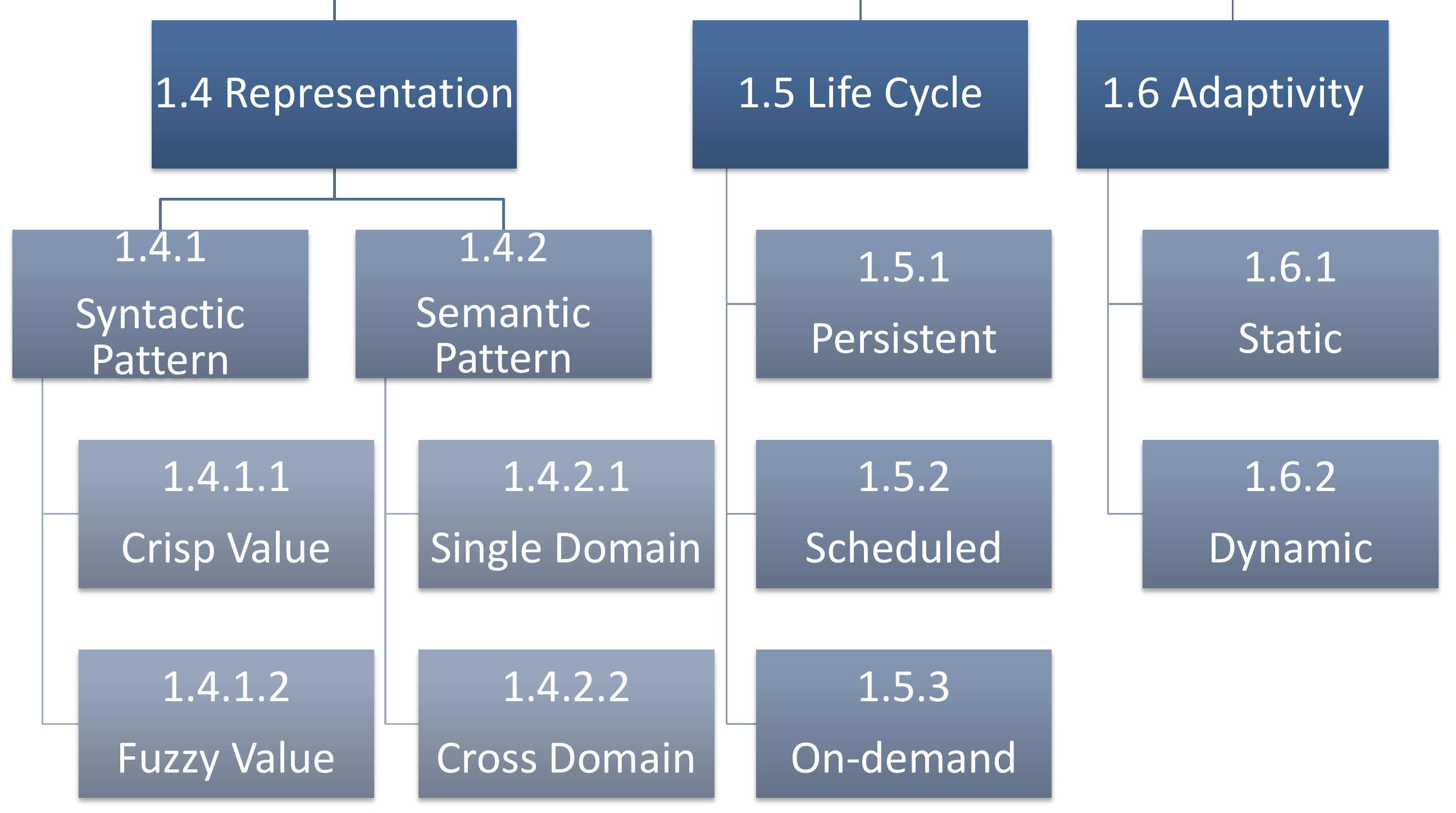}
  \caption{Representation, Life Cycle and Adaptivity dimensions}
  \label{fig:taxonomy4}
\end{figure}

As shown in Figure \ref{fig:taxonomy4}, DR patterns are specified at different abstraction levels,
primarily determined by the underlying event models. If only using traditional CEP systems,
syntactic patterns have
to be defined over raw data streams. This has been explored in other
literature~\cite{Asaf2006, Dunkel2009}. The event attributes can be either crisp
values or fuzzy concepts, depending on the uncertainty in matching. Our semantic CEP framework
allows users to intuitively define patterns over one or more domain
ontologies. Examples 1--6 illustrate such patterns.

%

%
%
%

\subsection{Life Cycle}
The life cycle of an event pattern is the time period during which it is active. As shown in Figure
\ref{fig:taxonomy4}, some DR patterns may run persistently, some only be active for scheduled
periods, and others activated on-demand (say by other patterns that are detected). Most
monitoring and prediction patterns are persistent. However, curtailment patterns are meaningful only
when there is a potential peak load, say, after receiving a DR request from the utility. Since
there is a resource cost associated with having patterns active, these patterns are active
on-schedule or on-demand.

\subsection{Adaptivity}
DR patterns may be categorized based on how often they evolve over time. Some patterns may be static
and do not need to change after they were firstly introduced. However, some patterns such as
prediction patterns may be affected by changes in the grid
infrastructure and consumer behavior. A novel area of research is to mine historical event
streams to automate the process of defining interesting patterns, allowing patterns to self-adapt.


\section{USC Microgrid Case Study}
\label{sec:casestudy}
The taxonomy was informed through DR approaches that were investigated in the USC campus
microgrid. Events representing different dimensions in the taxonomy were implemented and their
efficacy evaluated in the campus. We present those results here.  
We use \emph{SCEPter}, our semantic complex event processing engine \cite{zhou2012SCEPTechReport}, to
detect semantic CEP patterns defined over a selected set of event streams in the campus BAN. These
patterns span different DR end-uses: monitoring, prediction and curtailment, and these experiments
were conducted over a 4~day period on campus.

\subsection{Events and Ontologies}
USC microgrid event streams used in our experiments are:
\begin{itemize}
    \item \emph{Meter measurement.} Events from smart meters which measure buildings' KW loads.
    \item \emph{Fan coil status.} Events from HVAC sensors which report the operational status of fan
      coils: ``1'' means ON and ``0'' means OFF.
    \item \emph{Class schedule.} Data from a calendar schedule service which generates a classroom
      schedule event an hour before a class begins.
    \item \emph{Room temperature.} Measurement from room-level space temperature sensor.
    \item \emph{Room occupancy.} Events from room occupancy sensors that provides boolean readings.
\end{itemize}

Events from the same type of sources are pushed to a single logical stream. The campus microgrid
domain ontologies are described in \cite{zhou2012ITNG}. These capture properties of and relationships between physical space, electric equipments, and organizations on campus.

\subsection{DR Patterns and Empirical Evaluations}
\begin{figure}[!h!t!b]
  \centering
  \includegraphics[width=3.5in]{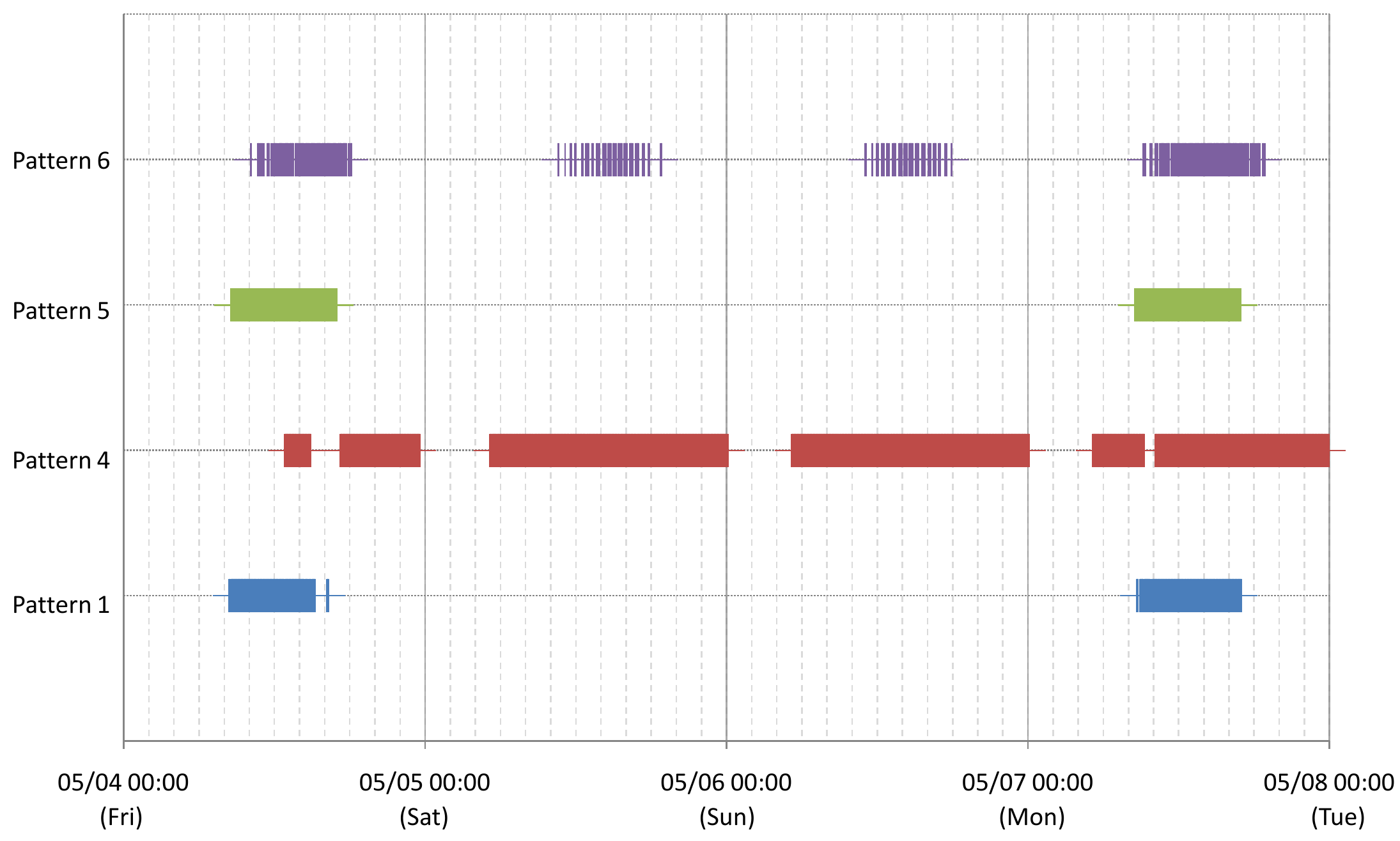}
  \caption{Experiment Results}
  \label{fig:exp}
\end{figure}

The four DR patterns introduced in Section \ref{sec:taxonomy} are evaluated over the above event
streams. Specifically, we analyze the detection of pattern \textbf{1} (average power consumption exceeds a
peak load), \textbf{4} (space temperature of unoccupied room less than 73$^\circ$F), \textbf{5} (more than six fan
coils are concurrently active) and \textbf{6} (load on EE department exceeds 600KW).


The experiments were conducted from Friday May 4$^{th}$ to Tuesday May 7$^{th}$, 2012. Figure
\ref{fig:exp} shows the detection of these four patterns over the six event streams during that time
period. The detection frequency of some patterns were limited since this time period coincided with
the final
exam week when classes and DR curtailment were not actively scheduled.

In Figure \ref{fig:exp}, 
pattern 1's detection indicates that the power consumption of the MHP building exceeded its pre-peak
threshold from around 8:20AM to 4:00PM on Friday and from around 8:40AM to 5:00PM on Monday. The power
load of MHP during weekends is below the pre-peak threshold because it is primarily used for teaching.
However, we observe from Pattern 6 that the power consumption of the EE department
exceeds its pre-peak threshold even on the weekend. Detection of these patterns helped the facility managers decide
when and where to curtail energy use on campus -- these patterns do not activate actual
curtailments yet, but offer an insight into the potential.

Pattern 4 and 5 show opportunities
for curtailments. From pattern 5, we know that more than 6 fan coils in MHP operate concurrently
from $\sim$8:00AM to 5:00PM on weekdays. By duty cycling the operations of fan coils during this
period, we can flatten the demand curve. In a separate experiment, we observed over 27\% curtailment in
peak demand by duty cycling fan coils in MHP. Pattern 4 monitors a meeting room
in the EE department. Several group meetings were scheduled on Friday and Monday. It is
observed that as people leave the room without resetting the thermostat, it causes power wastage
when the room is unoccupied -- which is during most times, especially the weekend.

These patterns and situations are detected in realtime, which helps undertake fine grained,
timely and intelligent DR strategies. We are currently
developing the action rule engine that can initiate automated actions and help complete the
event-based DR loop in Figure \ref{fig:ra}. A comprehensive suite of experiments
across $\sim$40 buildings on campus is planned for the next peak load season. This will offer
an accurate estimate of the improvement in curtailment response using dynamic event-based DR
approaches as compared to static schedules.  
\section{Related Work}
\label{sec:relatedwork}
Existing DR strategies use incentive-based and time-based programs. 
Incentive-based programs such as dynamic pricing offer benefits to customers who perform voluntary
curtailment. This requires manual intervention by customers and the outcome is less
reliable. Open Automated Demand Response Communications Specifications (OpenADR) model \cite{Thatikar2010, Mathieu2011} is
increasingly used to communicate pricing signals to customers in realtime. These signals are mapped to
operation modes of building control systems through production rules. Our work supplements this
approach by providing the capabilities to correlate heterogeneous microgrid events to initiate and
target the curtailment strategies.

Time-based demand schedules are commonly used for DR in Smart Grids. These approaches model DR as a
mathematical optimization problem, maximizing the user or the utility's benefit.
In \cite{ParvaniaSG2010}, optimal schedules of generation units and demand-side reserves were discussed, where the objective function was formulated as a two-stage stochastic programming model.
In \cite{RadSG2010A}, DR models were proposed for a single household which schedule appliance activities attempting to minimize user bills.
In \cite{Bagherian2009, Choi2011}, the authors discuss models for microgrid which compute the optimum energy plan, i.e., the
amount of power to be purchased, sold, transferred, and stored for a time period to
minimize the total operation cost.
Nevertheless, these DR approaches are predicated on accurate mathematical modeling which require
in-depth knowledge of the system and are not sustainable as the power grid evolves continuously, and
unpredictable events that influence power consumptions occur dynamically. An opportunistic DR scheme driven by real-time monitoring data can hence
supplement these existing approaches.

CEP itself has received much attention in a variety of domains \cite{Asaf2006, Dunkel2009}. There is also increasing interest using CEP for Smart Grid applications recently. In \cite{Renners2012}, the authors proposed a CEP approach to detect building occupancy changes for energy saving. However their event patterns are specified at low level over raw events. In \cite{Wagner2010}, the vision of using CEP over linked Smart Grid data was discussed in general, but these are anecdotal rather than comprehensive uses of CEP. In \cite{Stojanovic2011}, a semantic CEP system for light management in smart offices was introduced while the domain ontologies only capture spatial semantics of lightning devices. A comprehensive analysis of event patterns to guide event-based Smart Grid application development is still missing. To our knowledge, our work is among the first efforts to analyze and implement semantic CEP for DR applications at a microgrid scale.
\section{Conclusion}
\label{sec:conclusion}
We have discussed the use of semantic CEP for dynamic demand-side management in a
campus microgrid. By abstracting realtime sensor information and domain knowledge as semantic
events, our approach enables DR end-use needs, such as as monitoring, prediction and
curtailment, to be intuitively modeled as high-level patterns without knowledge of raw events. Our
taxonomy, informed and validated by DR techniques in the microgrid, offers a structure for
operators to develop their own suite of DR patterns for their service area.  We believe CEP
offers a powerful analytical tool for achieving realtime DR, but this requires in-depth study of
their real world use; our work is a step to translate this potential into reality. The
ability to automatically mine for self-adaptive patterns can lead to a
paradigm shift in informatics-driven demand management for a reliable and efficient Smart Grid.

\vspace{10pt}






%
\bibliographystyle{abbrv}
\bibliography{qunzhiRef}

\begin{IEEEbiography}[{\includegraphics[width=1in,height=1.25in,clip,keepaspectratio]{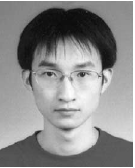}}]{Qunzhi Zhou}
is a PhD student in Computer Science Department at the University of Southern California and a Research Assistant in the Smart Grid project at the USC Center for Energy Informatics. His research interests are in the areas of complex event processing, semantic web, machine learning and distributed computing systems. He has a M.S. in Computer Science from the University of Southern California and received his B.S. in Automation from Tsinghua University, China. 
\end{IEEEbiography}
\vspace{-100mm}
\begin{IEEEbiography}[{\includegraphics[width=1in,height=1.25in,clip,keepaspectratio]{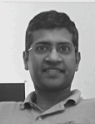}}]{Yogesh Simmhan} is a Research Assistant Professor in the Electrical Engineering Department and
Associate Director of the Center for Energy Informatics, University of Southern California. His research focuses
on distributed data and computing systems, workflows for dynamic applications, and scalable data management for eScience and eEngineering. He has a PhD in Computer Science from Indiana University and was earlier
with Microsoft Research. He is a IEEE and ACM member.
\end{IEEEbiography}
\vspace{-100mm}
\begin{IEEEbiography}[{\includegraphics[width=1in,height=1.25in,clip,keepaspectratio]{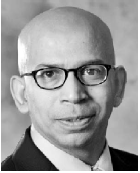}}]{Viktor Prasanna} is the Charles Lee Powell Chair in Engineering and is Professor of Electrical Engineering
and Professor of Computer Science at the University of Southern California. He serves as the Director of USC Center for Energy Informatics. His research interests include High Performance Computing, Parallel and Distributed Systems, Reconfigurable Computing, and Embedded Systems. He received his M.S. from the School of Automation, Indian Institute of Science and PhD in Computer Science from Pennsylvania State University. He is a Fellow of IEEE, ACM and AAAS.
\end{IEEEbiography}

\end{document}